# The potentials of the acceleration field and pressure field in rotating relativistic uniform system

**Sergey G. Fedosin**

PO box 614088, Sviazeva str. 22-79, Perm, Perm Krai, Russia

E-mail: fedosin@hotmail.com

**Abstract**: The scalar and vector potentials of the acceleration field and the pressure field are calculated for the first time for a rotating relativistic uniform system, and the dependence of the potentials on the angular velocity is found. These potentials are compared with the potentials for the non-rotating uniform system that have been found previously. The rotation leads to the appearance of vector potentials, which at each point turn out to be directed along the corresponding linear velocity of rotation. The calculation shows that for rotating stellar objects the contribution to the fields' vector potentials from the proper random motion of particles is small compared to the contribution from rotation and may not be taken into account. From the expression for the pressure field potential a relativistic formula follows that relates the pressure, mass density, and mean square velocity of the particles. This formula in the limit of low speeds corresponds to the expression for the pressure in molecular kinetic theory. When calculating the potentials, a new method is used that takes into account the potentials of two different bodies, a cylinder and a sphere, for solving the wave equation of rotating system.

## 1. Introduction

The derivation of formulas for the pressure field in a moving matter is not a trivial problem and in the general case requires the solution of differential equations. One of the methods is to solve the Navier-Stokes equation for a given velocity field and known boundary conditions [1]. The expression for the pressure in [2] was obtained by generalizing the Bernoulli equation, in [3-4] the pressure field was found numerically using equations containing pressure and a velocity field. In this case, the pressure field in the matter is considered simplified as a scalar field depending on time and coordinates. In [5], the pressure field, taking into account electromagnetic phenomena, is considered within the framework of hydrodynamics and thermodynamics, using the approaches of Euler and Lagrange when describing the motion of matter. There are also known works on determining the potentials of the electromagnetic field in rotating bodies [6-11].

The aim of this article is to study a rotating relativistic uniform system using the methods of analytical mechanics in the approximation of continuously distributed matter.

The relativistic uniform system is one of the models of modern physics and finds its application in various areas [12-16]. This model describes various gravitationally coupled systems based on field theory and therefore turns out to be much more accurate compared to the model of a uniform system of classical mechanics. The relativistic model generalizes the classical uniform system model and uses four-dimensional formalism instead of three-dimensional vectors. Unlike the phenomenological thermodynamic approach, the pressure field is treated not as a scalar, but as a vector field with its own four-potential, pressure tensor and energy-momentum tensor.

The field equations allow using the principle of least action to derive the relations arising between the forces acting in a matter. The smaller the density gradients of the matter inside the system, the closer in its properties this system becomes to a relativistic uniform system. As a result, the theoretical results obtained can be applied, for example, to the dynamics of cosmic gas clouds and to the Metagalaxy as a whole. It may be interesting to study the motion of magnetized matter in the Earth's core in connection with the problem of the origin of magnetic field and change of magnetic poles.

In [17-21] the potentials and strengths of various vector fields in a uniform system at rest were calculated. In such a system, the global vector field potentials are equal to zero. The next step is calculating the potentials for a rotating relativistic uniform system. The need for this step follows from the fact that only taking into account the vector potentials that appear during rotation allows us to correctly calculate such physical quantities as angular momentum and relativistic energy of system. It is well known that in the presence of gravitational and electromagnetic fields, instead of the usual momentum and angular momentum, generalized momenta are used, expressed through the vector potentials of the fields. These quantities can be used to estimate the corresponding quantities for the observed bodies, such as nucleons and neutron stars, since these objects are quite close in their properties to the rotating relativistic uniform system.

If we talk about the potentials of the acceleration field and the pressure field, then using these potentials it becomes possible to calculate the dependence of mass density on coordinates, the distribution of velocities of typical particles, pressure and temperature in the system under consideration. Previously, similar calculations were made without taking into account the rotation of a relativistic uniform system and were used to estimate the parameters of stars and planets [13], to calculate the kinetic energy of the system [14], to determine the dependence of the root mean square velocity on coordinates [16].

Further, we will assume that the physical system has a spherical shape and is so densely filled with particles that we can use the continuous medium approximation and calculate various quantities using integrals over the system's volume. The main fields are usually considered the gravitational and electromagnetic fields, acceleration field and pressure field, created by the particles, and these fields maintain both the system's integrity and the state of its constant rotation.

In this article, in order to reduce its volume, we will focus only on finding the potentials of the acceleration field and pressure field. The peculiarity of these fields has one more reason – usually they act locally, in the sense that they vanish outside the matter. The locality of action of the acceleration field and pressure field allows us to significantly simplify the solution of the equations for the potentials of these fields.

In contrast, the gravitational and electromagnetic fields can also exist outside the matter. Apparently, this is due to the fundamental difference between these fields. There is a theoretical approach, according to which the gravitons and charged particles of the electrogravitational vacuum act inside and outside the bodies, leading to the corresponding forces in the matter of these bodies [22-23]. In order to describe this action, the concepts of the gravitational and electromagnetic fields are introduced. The pressure field takes into account the forces that appear during mechanical collisions of the particles with each other. In order to describe the force of deceleration of the motion of the adjacent directed and interacting particles' fluxes, the dissipation field is introduced [12].

If we consider all these fields as vector fields, then the acceleration field that defines the particles' acceleration must also be assumed to be a vector field. Indeed, in the equation of the particles' motion the acceleration is associated with the acting forces, which are determined through field tensors [24]. In addition, taking into account the acceleration field allows one to obtain in covariant form not only acceleration, but also energy, including the rest energy of the system [15]. This becomes possible due to the fact that the acceleration field as an independent field is introduced into the Lagrange function. The subsequent application of the principle of least action makes it possible to derive the equations of the acceleration field and obtain the equation of motion [19]. In the simplest case, the acceleration field sets the inertial force in the form of the product of mass and acceleration, while other fields set the forces acting on a given mass. Then the connection between the force of inertia and the sum of forces from other fields is expressed by the well-known second Newton's law. In particular, the acceleration field is used to accurately reproduce the Navier-Stokes equation [12].

## 2. The acceleration field

The equations for the acceleration field resemble in their form Maxwell equations and are presented in [19]. Within the framework of the special theory of relativity they look as follows:

$$\nabla \cdot \mathbf{S} = 4\pi \eta \gamma \rho_0, \quad \nabla \times \mathbf{N} = \frac{1}{c^2}\frac{\partial \mathbf{S}}{\partial t} + \frac{4\pi \eta \gamma \rho_0 \mathbf{v}}{c^2}, \quad \nabla \cdot \mathbf{N} = 0, \quad \nabla \times \mathbf{S} = -\frac{\partial \mathbf{N}}{\partial t}, \quad (1)$$

where $\eta$ is the acceleration field coefficient, $\gamma = \dfrac{1}{\sqrt{1 - v^2/c^2}}$ is the Lorentz factor for the particles moving inside the sphere, $\mathbf{v}$ is the particles' velocity in the reference frame $K$ associated with the center of the sphere, $\rho_0$ is the mass density of an arbitrary particle in the comoving reference frame, $c$ is the speed of light, $\mathbf{S}$ and $\mathbf{N}$ are the strength and the solenoidal vector of the acceleration field, respectively.

The four-potential $U_\mu = \left(\dfrac{\vartheta}{c}, -\mathbf{U}\right)$ is defined with the help of the scalar potential $\vartheta$ and the vector potential $\mathbf{U}$ of the acceleration field in such a way that the following relations hold true:

$$\mathbf{S} = -\nabla \vartheta - \frac{\partial \mathbf{U}}{\partial t}, \qquad \mathbf{N} = \nabla \times \mathbf{U}. \qquad (2)$$

Substitution of (2) into (1) leads to the wave equations for the potentials:

$$\partial_\beta \partial^\beta \vartheta = \frac{1}{c^2}\frac{\partial^2 \vartheta}{\partial t^2} - \Delta \vartheta = 4\pi \eta \rho_0 \gamma, \qquad \partial_\beta \partial^\beta \mathbf{U} = \frac{1}{c^2}\frac{\partial^2 \mathbf{U}}{\partial t^2} - \Delta \mathbf{U} = \frac{4\pi \eta}{c^2}\mathbf{J}.$$

According to (1-2), the acceleration field is described by two potentials $\vartheta$ and $\mathbf{U}$, and two vectors $\mathbf{S}$ and $\mathbf{N}$. The vector $\mathbf{N}$ in its properties is similar to the magnetic field arising due to the movement of charged particles and due to the delay in the propagation of electromagnetic influence. It is known that for a stationary particle both its vector potential and its magnetic field are equal to zero. The same is true for the solenoidal vector $\mathbf{N}$ reflecting the relativistic nature of the acceleration field.

If we consider the average quantities, then in the stationary system the particles' velocity $\mathbf{v}$, the vector of the mass current density $\mathbf{J} = \gamma \rho_0 \mathbf{v}$, as well as the potentials $\vartheta$ and $\mathbf{U}$ do not depend on the time, but are functions of the space coordinates. As a result, only the Laplacians remain in the wave equations:

$$\Delta \vartheta = -4\pi \eta \rho_0 \gamma, \qquad \Delta \mathbf{U} = -\frac{4\pi \eta \rho_0 \gamma}{c^2} \mathbf{v}. \qquad (3)$$

In order to simplify, we will assume that the matter inside the sphere rotates with the same constant angular velocity $\omega$ relative to the axis that passes through the center of the sphere. Using the rule of relativistic addition of velocities, for the absolute velocity and the Lorentz factor of an arbitrary particle in the reference frame $K$, we can write the following:

$$\mathbf{v} = \frac{\mathbf{v}' + \frac{(\gamma_r - 1)(\mathbf{v}'\mathbf{v}_r)}{v_r^2}\mathbf{v}_r + \gamma_r \mathbf{v}_r}{\gamma_r \left(1 + \frac{\mathbf{v}'\mathbf{v}_r}{c^2}\right)}, \qquad \gamma = \gamma' \gamma_r \left(1 + \frac{\mathbf{v}'\mathbf{v}_r}{c^2}\right), \qquad (4)$$

where $\mathbf{v}'$ is the velocity of the random motion of the particle in the reference frame $K'$ that rotates with the matter at the angular velocity $\omega$; $\mathbf{v}_r$ is the linear velocity of motion of the reference frame $K'$ at the location of the particle, resulting from rotation in the reference frame $K$; $\gamma_r = \frac{1}{\sqrt{1 - v_r^2/c^2}}$ is the Lorentz factor for the velocity $\mathbf{v}_r$, $\gamma' = \frac{1}{\sqrt{1 - v'^2/c^2}}$ is the Lorentz factor for the velocity $\mathbf{v}'$.

Let us average expressions (4) over the volume in a small neighborhood around the point under consideration, so that a sufficient number of particles would be present in this volume. The velocity of random motion $\mathbf{v}'$ is presented in (4) in such a way that, after averaging, the contributions from the terms containing $\mathbf{v}'$ become equal to zero. This happens because the velocity $\mathbf{v}'$ of different particles has a different direction, including the opposite. Then it follows from (4) for the average quantities: $\bar{\mathbf{v}} = \mathbf{v}_r$, $\bar{\gamma} = \gamma' \gamma_r$.

For the spherical system with the particles in the absence of the matter's general rotation, the Lorentz factor depends on the current radius $r$ and on the value of the Lorentz factor $\gamma'_c$ at the center of the sphere [17]:

$$\gamma'(\omega=0) = \frac{c\gamma'_c}{r\sqrt{4\pi\eta\rho_0}}\sin\left(\frac{r}{c}\sqrt{4\pi\eta\rho_0}\right) \approx \gamma'_c - \frac{2\pi\eta\rho_0 r^2 \gamma'_c}{3c^2}.$$

This expression was obtained under the assumption that the matter's particles do not have their proper rotation and move inertially and randomly. We will further assume that in the reference frame $K'$, rotating together with the matter at the angular velocity $\omega$, the Lorentz factor $\gamma'$ is determined by a similar formula

$$\gamma' = \frac{c\gamma_c}{r\sqrt{4\pi\eta\rho_0}}\sin\left(\frac{r}{c}\sqrt{4\pi\eta\rho_0}\right) \approx \gamma_c - \frac{2\pi\eta\rho_0 r^2 \gamma_c}{3c^2}. \qquad (5)$$

with the exception that the Lorentz factor $\gamma_c$ is used at the center of the rotating sphere, which may differ in magnitude from $\gamma'_c$ for the sphere at rest.

This means that if in the reference frame $K$ there is a rotating relativistic uniform system, then when turning to the comoving rotating reference frame $K'$ (5) would hold true and the general rotation would not change the dependence of the Lorentz factor $\gamma'$ on the current radius.

In its meaning, such a situation corresponds to the principle of relativity, when in an inertial reference frame a simultaneous increase of the velocity of all the system's particles by the same value does not change the state of the system. The only difference is that the rotation changes the particles' velocities differently depending on the current radius, which makes the reference frame non-inertial. Therefore, the assumption of application of (5) for the rotating reference frame $K'$ can be considered as the first approximation to reality.

In spherical coordinates $r, \theta, \varphi$ the amplitude of the linear rotation velocity is expressed by the formula: $v_r = \omega r \sin\theta$, where the angle $\theta$ is measured from the axis $OZ$, and the matter's particles rotate in the planes parallel to the plane $XOY$. In cylindrical coordinates $\rho, \varphi, z$ the velocity is given by the expression $v_r = \omega\rho$.

Consequently, the Lorentz factor can be written as follows:

$$\gamma_r = \frac{1}{\sqrt{1-v_r^2/c^2}} = \frac{1}{\sqrt{1-\omega^2 r^2 \sin^2\theta/c^2}} = \frac{1}{\sqrt{1-\omega^2 \rho^2/c^2}}, \qquad (6)$$

where $\rho$ is the distance from the axis of rotation to the point under consideration.

Instead of the real matter's particles, it is convenient to assume that the matter consists of typical particles, which characterize on the average all the system's properties. After averaging of all the quantities over the volume, it turns out that the equations for typical particles relate the averaged physical quantities with each other.

Substitution of $\bar{\gamma} = \gamma'\gamma_r$ instead of $\gamma$ and $\bar{\mathbf{v}} = \mathbf{v}_r$ instead of $\mathbf{v}$ in (3), taking into account (5) and (6) gives the following:

$$\Delta\vartheta = -4\pi\eta\rho_0\gamma'\gamma_r = -\frac{c\gamma_c\sqrt{4\pi\eta\rho_0}}{r\sqrt{1-\omega^2 r^2 \sin^2\theta/c^2}}\sin\left(\frac{r}{c}\sqrt{4\pi\eta\rho_0}\right). \qquad (7)$$

$$\Delta\mathbf{U} = -\frac{\gamma_c\sqrt{4\pi\eta\rho_0}}{cr\sqrt{1-\omega^2 r^2 \sin^2\theta/c^2}}\sin\left(\frac{r}{c}\sqrt{4\pi\eta\rho_0}\right)\mathbf{v}_r. \qquad (8)$$

In spherical coordinates for the arbitrary function $f = f(r,\theta,\varphi)$ the Laplacian of this function for the case $r \neq 0$ has the form:

$$\Delta f = \frac{1}{r}\frac{\partial^2}{\partial r^2}(rf) + \frac{1}{r^2\sin\theta}\frac{\partial}{\partial\theta}\left(\sin\theta\frac{\partial f}{\partial\theta}\right) + \frac{1}{r^2\sin^2\theta}\frac{\partial^2 f}{\partial\varphi^2}. \qquad (9)$$

For the sphere at rest at $\omega = 0$, equation (3) for the scalar potential and the solution of this equation have the form:

$$\Delta\vartheta' = -4\pi\eta\rho_0\gamma'(\omega=0), \qquad \vartheta' = c^2\gamma'(\omega=0) = \frac{c^3\gamma'_c}{r\sqrt{4\pi\eta\rho_0}}\sin\left(\frac{r}{c}\sqrt{4\pi\eta\rho_0}\right). \qquad (10)$$

The potential $\vartheta'$ in (10) depends only on the radius $r$. Substitution of $\vartheta'$ instead of $f$ in (9) allows us to calculate the Laplacian $\Delta\vartheta'$ and compare the result with the first expression in (10).

Let us solve an auxiliary problem and find the scalar potential $\vartheta_r$ of the acceleration field for the case of purely rotational motion of uniform system in the form of an infinite long cylinder in the absence of random motion of the particles. From (3) and (6) it follows:

$$\Delta\vartheta_r = -4\pi\eta\rho_0\gamma_r = -\frac{4\pi\eta\rho_0}{\sqrt{1-v_r^2/c^2}} = -\frac{4\pi\eta\rho_0}{\sqrt{1-\omega^2\rho^2/c^2}}. \tag{11}$$

The Laplacian of the arbitrary function $f = f(\rho,\varphi,z)$ in cylindrical coordinates has the following form:

$$\Delta f = \frac{1}{\rho}\frac{\partial}{\partial\rho}\left(\rho\frac{\partial f}{\partial\rho}\right) + \frac{1}{\rho^2}\frac{\partial^2 f}{\partial\varphi^2} + \frac{\partial^2 f}{\partial z^2}. \tag{12}$$

After replacing $f$ with $\vartheta_r$ in (12), taking into account (11), we arrive at the equation:

$$\frac{1}{\rho}\frac{\partial}{\partial\rho}\left(\rho\frac{\partial\vartheta_r}{\partial\rho}\right) = -\frac{4\pi\eta\rho_0}{\sqrt{1-\omega^2\rho^2/c^2}}. \tag{13}$$

In (13) the potential $\vartheta_r$ does not depend on the angle $\varphi$ due to symmetry with respect to the axis of rotation. It is also assumed that $\vartheta_r$ also does not depend on the coordinate $z$ inside the cylinder. This corresponds to the fact that $\vartheta_r$ is some function of the Lorentz factor (6) of the particles of the rotating matter, which depends only on the coordinate $\rho$ and the angular velocity of rotation $\omega$. It should be noted that when a free particle or particles of matter of a relativistic uniform system move, the acceleration field potential is simply proportional to the corresponding Lorentz factor. The fact that $\vartheta_r$ depends only on $\rho$ is the simplest choice for the dependence of the potential of the rotating cylinder. We will consider $\vartheta_r$ as an auxiliary potential for determining the total potential of the rotating spherical uniform relativistic system.

The solution of equation (13) is as follows:

$$\vartheta_r = \frac{4\pi \eta \rho_0 c^2}{\omega^2}\sqrt{1-\omega^2 \rho^2/c^2} - \frac{4\pi \eta \rho_0 c^2}{\omega^2}\ln\frac{1+\sqrt{1-\omega^2 \rho^2/c^2}}{\omega \rho/c} + C_1 \ln \rho + C_2.$$

The constants $C_1$ and $C_2$ must be chosen in such a way that $\vartheta_r$ must not depend either on $\ln(\omega \rho/c)$ nor on $\ln \rho$, which extend to infinity at $\rho = 0$. This gives the following:

$$\vartheta_r = \frac{4\pi \eta \rho_0 c^2}{\omega^2}\sqrt{1-\omega^2 \rho^2/c^2} - \frac{4\pi \eta \rho_0 c^2}{\omega^2}\ln\left(1+\sqrt{1-\omega^2 \rho^2/c^2}\right) + C_3. \qquad (14)$$

The question of choosing the constant $C_3$ will be solved in (20), after solving equation (7) for the rotating spherical uniform relativistic system.

In the case of rectilinear motion of a solid material point without its proper rotation, its scalar potential is equal to $\vartheta_p = \gamma_p c^2$, where $\gamma_p$ is the Lorentz factor of the particle. As for the vector potential of the acceleration field for such a solid point, it is equal to $\mathbf{U}_p = \gamma_p \mathbf{v}_p$, where $\mathbf{v}_p$ is the particle's velocity [25].

In this connection, we will multiply $\bar{\gamma} = \gamma' \gamma_r$ by $c^2$ and will consider the quantity $\vartheta^* = \gamma' \gamma_r c^2$ as the first approximation to the sought value of the scalar potential $\vartheta$. Taking into account that $x^2 + y^2 = \rho^2$ and $x^2 + y^2 + z^2 = r^2$, we will express $\gamma'$ in (5) and $\gamma_r$ in (6) in terms of the coordinates $x, y, z$, and will calculate the Laplacian $\Delta \vartheta^*$ in Cartesian coordinates:

$$\Delta \vartheta^* = \Delta \left[ \frac{c^3 \gamma_c}{\sqrt{4\pi \eta \rho_0}} \frac{1}{\sqrt{x^2+y^2+z^2}} \frac{1}{\sqrt{1-\omega^2(x^2+y^2)/c^2}} \sin\left( \frac{\sqrt{4\pi \eta \rho_0}}{c}\sqrt{x^2+y^2+z^2} \right) \right] =$$

$$= -\frac{c\gamma_c \sqrt{4\pi \eta \rho_0}}{r\sqrt{1-\omega^2 \rho^2/c^2}} \sin\left(\frac{r}{c}\sqrt{4\pi \eta \rho_0}\right) + \frac{2c\gamma_c \omega^2\left(1+\dfrac{\omega^2 \rho^2}{2c^2}\right)}{r\sqrt{4\pi \eta \rho_0}\left(1-\omega^2 \rho^2/c^2\right)^{5/2}} \sin\left(\frac{r}{c}\sqrt{4\pi \eta \rho_0}\right) +$$

$$+ \frac{2\gamma_c \omega^2 \rho^2}{r^2\left(1-\omega^2 \rho^2/c^2\right)^{3/2}} \cos\left(\frac{r}{c}\sqrt{4\pi \eta \rho_0}\right) - \frac{2c\gamma_c \omega^2 \rho^2}{r^3\sqrt{4\pi \eta \rho_0}\left(1-\omega^2 \rho^2/c^2\right)^{3/2}} \sin\left(\frac{r}{c}\sqrt{4\pi \eta \rho_0}\right).$$

(15)

The result of calculating $\Delta \vartheta^*$ on the right-hand side of (15) is expressed in terms of the radius $r$ and the cylindrical coordinate $\rho$. Using the expansion of functions of the form $\cos x \approx 1 - \dfrac{x^2}{2} + ...,\ \sin x \approx x - \dfrac{x^3}{6} + ...,$ we will expand the cosine and sine in the last two terms on the right-hand side of (15). In this case, the main expansion terms will be cancelled out, and there will remain a term containing the square of the speed of light $c^2$ in the denominator:

$$\frac{2\gamma_c \omega^2 \rho^2}{r^2 (1-\omega^2 \rho^2/c^2)^{3/2}} \cos\left(\frac{r}{c}\sqrt{4\pi\eta\rho_0}\right) - \frac{2c\gamma_c \omega^2 \rho^2}{r^3 \sqrt{4\pi\eta\rho_0}(1-\omega^2 \rho^2/c^2)^{3/2}} \sin\left(\frac{r}{c}\sqrt{4\pi\eta\rho_0}\right) \approx$$

$$\approx -\frac{8\pi\eta\rho_0 \gamma_c \omega^2 \rho^2}{3c^2}.$$

We transform the second term on the right-hand side of (15), assuming that $\sin\left(\dfrac{r}{c}\sqrt{4\pi\eta\rho_0}\right) \approx \dfrac{r}{c}\sqrt{4\pi\eta\rho_0}$:

$$\frac{2c\gamma_c \omega^2 \left(1 + \dfrac{\omega^2 \rho^2}{2c^2}\right)}{r\sqrt{4\pi\eta\rho_0}(1-\omega^2 \rho^2/c^2)^{5/2}} \sin\left(\frac{r}{c}\sqrt{4\pi\eta\rho_0}\right) \approx \frac{2c\gamma_c \omega^2 \left(1 + \dfrac{3\omega^2 \rho^2}{c^2}\right)}{r\sqrt{4\pi\eta\rho_0}} \sin\left(\frac{r}{c}\sqrt{4\pi\eta\rho_0}\right) \approx$$

$$\approx \frac{2c\gamma_c \omega^2}{r\sqrt{4\pi\eta\rho_0}} \sin\left(\frac{r}{c}\sqrt{4\pi\eta\rho_0}\right) + \frac{6\gamma_c \omega^4 \rho^2}{c^2}.$$

Then, as the first approximation, we can write:

$$\Delta\vartheta^* \approx -\frac{c\gamma_c \sqrt{4\pi\eta\rho_0}}{r\sqrt{1-\omega^2 \rho^2/c^2}} \sin\left(\frac{r}{c}\sqrt{4\pi\eta\rho_0}\right) + \frac{2c\gamma_c \omega^2}{r\sqrt{4\pi\eta\rho_0}} \sin\left(\frac{r}{c}\sqrt{4\pi\eta\rho_0}\right) + \tag{16}$$

$$+ \frac{6\gamma_c \omega^4 \rho^2}{c^2} - \frac{8\pi\eta\rho_0 \gamma_c \omega^2 \rho^2}{3c^2}.$$

Now we will find such a function $\vartheta_1$ that its Laplacian would be equal to the sum of the last three terms on the right-hand side of (16):

$$\Delta\vartheta_1 = \frac{2c\gamma_c \omega^2}{r\sqrt{4\pi\eta\rho_0}} \sin\left(\frac{r}{c}\sqrt{4\pi\eta\rho_0}\right) + \frac{6\gamma_c \omega^4 \rho^2}{c^2} - \frac{8\pi\eta\rho_0 \gamma_c \omega^2 \rho^2}{3c^2}.$$

Taking into account the form of the Laplacian (9) in spherical coordinates, and, accordingly, the form of the Laplacian (12) in cylindrical coordinates, we find a solution that does not contain infinities at $r = 0$ and at $\rho = 0$:

$$\vartheta_1 = -\frac{\gamma_c \omega^2 c^3}{2\pi \eta \rho_0 r \sqrt{4\pi \eta \rho_0}} \sin\left(\frac{r}{c}\sqrt{4\pi \eta \rho_0}\right) + \frac{3\gamma_c \omega^4 \rho^4}{8c^2} - \frac{\pi \eta \rho_0 \gamma_c \omega^2 \rho^4}{6c^2} + C_4. \quad (17)$$

Subtracting term by term the equality for $\Delta \vartheta_1$ from equality (16), we obtain:

$$\Delta(\vartheta^* - \vartheta_1) \approx -\frac{c\gamma_c \sqrt{4\pi \eta \rho_0}}{r\sqrt{1-\omega^2 \rho^2/c^2}} \sin\left(\frac{r}{c}\sqrt{4\pi \eta \rho_0}\right).$$

From comparison of this equality with (7), taking into account $\vartheta^* = \gamma' \gamma_r c^2$ and equalities (5), (6), (17), we obtain an approximate expression for the scalar potential of the acceleration field:

$$\vartheta \approx \vartheta^* - \vartheta_1 = \frac{c^3 \gamma_c}{r\sqrt{4\pi \eta \rho_0}}\left(\frac{1}{\sqrt{1-\omega^2 \rho^2/c^2}} + \frac{\omega^2}{2\pi \eta \rho_0}\right)\sin\left(\frac{r}{c}\sqrt{4\pi \eta \rho_0}\right) - \\ -\frac{3\gamma_c \omega^4 \rho^4}{8c^2} + \frac{\pi \eta \rho_0 \gamma_c \omega^2 \rho^4}{6c^2} - C_4. \quad (18)$$

We will consider the inner region inside the sphere at small $r$ and $\rho$, that is, the region near the axis of rotation and near the center of the sphere. In this case, the sine in (18) can be replaced by its argument:

$$\vartheta(r \approx 0) \approx c^2 \gamma_c \left(\frac{1}{\sqrt{1-\omega^2 \rho^2/c^2}} + \frac{\omega^2}{2\pi \eta \rho_0}\right) - \frac{3\gamma_c \omega^4 \rho^4}{8c^2} + \frac{\pi \eta \rho_0 \gamma_c \omega^2 \rho^4}{6c^2} - C_4. \quad (19)$$

Using the freedom to calibrate the scalar potential, let us equate the potential of the rotating cylinder near the axis of the cylinder and the potential near the center of the rotating sphere. For this purpose, let us send the cylindrical coordinate $\rho$ in equalities (14) and in (19) approach zero and equate these equalities. This allows us to estimate the constant $C_3$ and specify the form of the scalar potential $\vartheta_r$ in (14):

$$C_3 = -\frac{4\pi\eta\rho_0 c^2}{\omega^2} + \frac{4\pi\eta\rho_0 c^2}{\omega^2}\ln 2 + c^2\gamma_c\left(1 + \frac{\omega^2}{2\pi\eta\rho_0}\right) - C_4.$$

$$\vartheta_r = \frac{4\pi\eta\rho_0 c^2}{\omega^2}\left[\sqrt{1-\omega^2\rho^2/c^2} - \ln\left(1+\sqrt{1-\omega^2\rho^2/c^2}\right) - 1 + \ln 2\right] + c^2\gamma_c\left(1 + \frac{\omega^2}{2\pi\eta\rho_0}\right) - C_4.$$

Hence it follows that on the axis of rotation, where $\rho = 0$, the potential is $\vartheta_r(\rho=0) = c^2\gamma_c\left(1 + \frac{\omega^2}{2\pi\eta\rho_0}\right) - C_4$. If we put the constant $C_4 = \frac{c^2\gamma_c\omega^2}{2\pi\eta\rho_0}$, then the potential on the axis of rotation of the cylinder will not depend on the angular velocity of rotation $\omega$. In this case, we find:

$$\vartheta_r = \frac{4\pi\eta\rho_0 c^2}{\omega^2}\left[\sqrt{1-\omega^2\rho^2/c^2} - \ln\left(1+\sqrt{1-\omega^2\rho^2/c^2}\right) - 1 + \ln 2\right] + c^2\gamma_c. \qquad (20)$$

At small $\omega$, the potential in (20) ceases to depend on $\omega$ and becomes approximately equal to the value $\vartheta_r(\omega \approx 0) \approx -\pi\eta\rho_0\rho^2 + c^2\gamma_c$, and there is a Lorentz factor $\gamma_c$ at the center of the sphere and at the center of the corresponding cylinder. To check this formula, it is sufficient to solve equation (13) at $\omega = 0$ for the case of a stationary cylinder. In this case, the solution is:

$$\vartheta_r(\omega = 0) = -\pi\eta\rho_0\rho^2 + C_5\ln\rho + C_6.$$

In the solution, it is necessary to set $C_5 = 0$ in order to avoid infinity at $\rho = 0$, and take into account that at $\rho = 0$ the potential must be equal to $c^2\gamma_c$. Therefore, there is $C_6 = c^2\gamma_c$, and we come to that $\vartheta_r(\omega = 0) = -\pi\eta\rho_0\rho^2 + c^2\gamma_c$.

Expression (20) has an interesting analogy with the dependence of chemical potential of a rotating solid in a state of thermal equilibrium [26], which has the following form in our notation:

$$\mu = u(x,y,z) + \mu_0(p,T) = -\frac{m\omega^2 \rho^2}{2} + \mu_0(p,T),$$

where $\mu_0(p,T)$ is a chemical potential in the absence of rotation, depending on pressure $p$ and temperature $T$; $m$ is the mass of a molecule; $u(x,y,z)$ in the general case is the potential energy of a molecule in a certain field, this energy for the case of a field of rotation was equated to the value of kinetic energy $\frac{m\omega^2 \rho^2}{2}$ of the molecule taken with a negative sign.

As the coordinate $\rho$ grows and moves away from the rotation axis, the chemical potential decreases, and the same follows for dependence of the potential $\vartheta_r$ in (20). In this case, the potential $c^2 \gamma_c$ can be considered an analogue for $\mu_0(p,T)$, and the first term on the right-hand side of (20) an analogue for $u(x,y,z)$.

Substitution $C_4 = \dfrac{c^2 \gamma_c \omega^2}{2\pi \eta \rho_0}$ in (18) gives:

$$\vartheta \approx \frac{c^3 \gamma_c}{r\sqrt{4\pi \eta \rho_0}} \left( \frac{1}{\sqrt{1-\omega^2 \rho^2/c^2}} + \frac{\omega^2}{2\pi \eta \rho_0} \right) \sin\left( \frac{r}{c}\sqrt{4\pi \eta \rho_0} \right) - \\ -\frac{3\gamma_c \omega^4 \rho^4}{8c^2} + \frac{\pi \eta \rho_0 \gamma_c \omega^2 \rho^4}{6c^2} - \frac{c^2 \gamma_c \omega^2}{2\pi \eta \rho_0}. \tag{21}$$

If the angular velocity of rotation $\omega$ vanishes in (21), the potential $\vartheta$ becomes equal to the potential $\vartheta'$ in (10).

### 3. The vector potential of the acceleration field

In the relativistic uniform system, which is generally at rest, the vector potentials of the fields, including the acceleration field, are equal to zero due to the particles' random motion. This is due to the superposition principle, when the vector potentials of individual particles are summed in a vector way.

For the case of rotation of the cylinder without taking into account the proper motion of the matter's particles from (3) and (6) in cylindrical coordinates, we obtain the following equation for the vector potential $\mathbf{U}_r$:

$$\Delta \mathbf{U}_r = -\frac{4\pi \eta \rho_0 \gamma_r}{c^2} \mathbf{v}_r = -\frac{4\pi \eta \rho_0}{c^2 \sqrt{1-\omega^2 \rho^2/c^2}} \mathbf{v}_r. \qquad (22)$$

From the relationship between Cartesian and cylindrical coordinates in the form $x = \rho\cos\varphi$, $y = \rho\sin\varphi$, $z = z$, it follows that in the absence of motion of the matter's particles in the direction of the axis $OZ$ the components of the linear velocity are determined as follows:

$$\mathbf{v}_r = \left(\frac{dx}{dt}, \frac{dy}{dt}, \frac{dz}{dt}\right) = (-\omega\rho\sin\varphi, \omega\rho\cos\varphi, 0) = (-\omega y, \omega x, 0). \qquad (23)$$

In (23) the angular velocity of rotation $\omega = \dfrac{d\varphi}{dt}$ is present.

We will project equation (22) on the axis $OX$ taking into account (23) and expression (12) for the Laplacian in cylindrical coordinates:

$$\Delta U_{rx} = \frac{1}{\rho}\frac{\partial}{\partial \rho}\left(\rho \frac{\partial U_{rx}}{\partial \rho}\right) + \frac{1}{\rho^2}\frac{\partial^2 U_{rx}}{\partial \varphi^2} = \frac{4\pi \eta \rho_0 \omega \rho \sin\varphi}{c^2 \sqrt{1-\omega^2 \rho^2/c^2}}. \qquad (24)$$

We will substitute into (24) the expression $U_{rx} = \rho Z \sin\varphi$, where the function $Z$ depends only on $\rho$. This allows us to get rid of $\sin\varphi$:

$$\frac{\partial}{\partial \rho}\left(\rho \frac{\partial(\rho Z)}{\partial \rho}\right) - Z = \rho^2 Z'' + 3\rho Z' = \frac{4\pi \eta \rho_0 \omega \rho^2}{c^2 \sqrt{1-\omega^2 \rho^2/c^2}}.$$

$$\rho Z'' + 3 Z' = \frac{4\pi \eta \rho_0 \omega \rho}{c^2 \sqrt{1-\omega^2 \rho^2/c^2}}.$$

Here $Z' = \dfrac{dZ}{d\rho}$, $Z'' = \dfrac{d^2 Z}{d\rho^2}$. The differential equation for calculating $Z$ can be transformed by multiplying by $d\rho$ and then integrating the right-hand side:

$$\rho dZ' = \frac{4\pi \eta \rho_0 \omega \rho}{c^2 \sqrt{1-\omega^2 \rho^2/c^2}} d\rho - 3 dZ.$$

$$\int \rho dZ' = \frac{4\pi\eta\rho_0\omega}{c^2}\int \frac{\rho}{\sqrt{1-\omega^2\rho^2/c^2}}d\rho - 3Z + C_7 = -\frac{4\pi\eta\rho_0}{\omega}\sqrt{1-\omega^2\rho^2/c^2} - 3Z + C_7.$$

The quantity $\rho dZ'$ can also be integrated by parts:

$$\int \rho dZ' = \rho Z' - \int Z'd\rho + C_8 = \rho Z' - Z + C_8.$$

Comparing the integrals for $\rho dZ'$ in the last two expressions, we find:

$$\rho Z' + 2Z = -\frac{4\pi\eta\rho_0}{\omega}\sqrt{1-\omega^2\rho^2/c^2} + C_9.$$

This equation has the following general solution:

$$Z = \frac{4\pi\eta\rho_0 c^2}{3\omega^3\rho^2}\left(1-\omega^2\rho^2/c^2\right)^{3/2} + \frac{C_{10}}{\rho^2} + \frac{C_9}{2}.$$

Now we can find the first component of the vector potential:

$$U_{rx} = \rho Z\sin\varphi = \left[\frac{4\pi\eta\rho_0 c^2}{3\omega^3\rho}\left(1-\omega^2\rho^2/c^2\right)^{3/2} + \frac{C_{10}}{\rho} + \frac{C_9}{2}\rho\right]\sin\varphi. \qquad (25)$$

In (25) it is still necessary to determine more precisely the values of the constant coefficients $C_9$ and $C_{10}$. By analogy with gauge-fixing of the scalar potential $\vartheta_r$ in (14), we will assume that if $\rho$ tends to zero, then the potential $U_{rx}$ must not extend to infinity. For this purpose, it suffices to choose the value of the constant $C_{10} = -\frac{4\pi\eta\rho_0 c^2}{3\omega^3}$. This gives:

$$U_{rx} = \left[\frac{4\pi\eta\rho_0 c^2}{3\omega^3\rho}\left(1-\omega^2\rho^2/c^2\right)^{3/2} - \frac{4\pi\eta\rho_0 c^2}{3\omega^3\rho} + \frac{C_9}{2}\rho\right]\sin\varphi. \qquad (26)$$

At low angular velocities $\omega$ the vector potential $\mathbf{U}_r$ must be proportional to the linear velocity $\mathbf{v}_r$ and the Lorentz factor $\gamma_r$, as is the case for the motion of the solid point: $\mathbf{U}_r \approx \gamma_r \mathbf{v}_r$. For the component $U_{rx}$, on condition that $\omega \to 0$, in view of (23), (26) and the expression in cylindrical coordinates $\gamma_r = \dfrac{1}{\sqrt{1 - \omega^2 \rho^2 / c^2}}$, according to (6), this can be written as follows:

$$\left[ \frac{4\pi \eta \rho_0 c^2}{3\omega^3 \rho}\left(1 - \omega^2 \rho^2/c^2\right)^{3/2} - \frac{4\pi \eta \rho_0 c^2}{3\omega^3 \rho} + \frac{C_9}{2}\rho \right]\sin\varphi \approx -\frac{\omega \rho \sin\varphi}{\sqrt{1 - \omega^2 \rho^2/c^2}}.$$

Hence it follows that $C_9 = \dfrac{4\pi \eta \rho_0}{\omega} - 2\omega$, and for projection of the vector potential on the axis $OX$ we obtain the following:

$$U_{rx} = \left[ \frac{4\pi \eta \rho_0 c^2}{3\omega^3 \rho}\left(1 - \omega^2 \rho^2/c^2\right)^{3/2} - \frac{4\pi \eta \rho_0 c^2}{3\omega^3 \rho} + \frac{2\pi \eta \rho_0 \rho}{\omega} - \omega \rho \right]\sin\varphi. \qquad (27)$$

Repeating all the steps, starting from (24), and projecting equation (22) on the axes $OY$ and $OZ$, we find the remaining components of the vector potential, respectively:

$$U_{ry} = -\left[ \frac{4\pi \eta \rho_0 c^2}{3\omega^3 \rho}\left(1 - \omega^2 \rho^2/c^2\right)^{3/2} - \frac{4\pi \eta \rho_0 c^2}{3\omega^3 \rho} + \frac{2\pi \eta \rho_0 \rho}{\omega} - \omega \rho \right]\cos\varphi, \qquad U_{rz} = 0.$$

$$(28)$$

From comparison of (23), (27), and (28) it follows that for the vector potential the following formula holds true:

$$\mathbf{U}_r = \left[ \frac{4\pi \eta \rho_0 c^2}{3\omega^4 \rho^2} - \frac{4\pi \eta \rho_0 c^2}{3\omega^4 \rho^2}\left(1 - \omega^2 \rho^2/c^2\right)^{3/2} - \frac{2\pi \eta \rho_0}{\omega^2} + 1 \right]\mathbf{v}_r. \qquad (29)$$

According to (29), the vector potential $\mathbf{U}_r$ is directed along the linear velocity of rotation $\mathbf{v}_r$ of the matter of the cylinder. Near the axis of rotation, where $\rho \approx 0$, and also at small values of the angular velocity $\omega$, the potential $\mathbf{U}_r$ in its value tends to the velocity of rotation $\mathbf{v}_r$.

In order to take into account the contribution to the vector potential from the particles' proper motion within the framework of the spheric relativistic uniform system, we will compare equation (8) for the potential $\mathbf{U}$ and equation (22) for the potential $\mathbf{U}_r$. The right-hand side of (8) turns out to be $\gamma'$ times greater than the right-hand side of (22), where $\gamma'$ is the Lorentz factor (5). In the first approximation, it should be assumed that the potential $\mathbf{U}$ will be $\gamma'$ times greater than the potential $\mathbf{U}_r$. Since $\gamma'$ is a function of the current radius $r$, we will determine the average value $\hat{\gamma}'$ as the average over the sphere's volume $V_s = \dfrac{4\pi a^3}{3}$, using in this case the volume element $dV = 4\pi r^2 dr$:

$$\hat{\gamma}' = \frac{1}{V_s} \int \gamma' dV = \frac{3c^2 \gamma_c}{4\pi \eta \rho_0 a^3} \left[ \frac{c}{\sqrt{4\pi \eta \rho_0}} \sin\left(\frac{a}{c}\sqrt{4\pi \eta \rho_0}\right) - a\cos\left(\frac{a}{c}\sqrt{4\pi \eta \rho_0}\right) \right]. \quad (30)$$

In most cases $\dfrac{a}{c}\sqrt{4\pi \eta \rho_0} \ll 1$, and then $\hat{\gamma}' \approx \gamma_c \left(1 - \dfrac{3\eta m}{10 a c^2}\right)$, where $m$ is the product of the mass density $\rho_0$ by the sphere's volume $V_s = \dfrac{4\pi a^3}{3}$, $\gamma_c$ denotes the Lorentz factor of the particles at the center of the sphere, and $a$ is the radius of the sphere. Multiplying $\mathbf{U}_r$ in (29) by $\hat{\gamma}'$ (30), we obtain the estimate of the vector potential $\mathbf{U}$ of the acceleration field for the rotating relativistic uniform system:

$$\mathbf{U} \approx \hat{\gamma}' \mathbf{U}_r = \hat{\gamma}' \left[ \frac{4\pi \eta \rho_0 c^2}{3\omega^4 \rho^2} - \frac{4\pi \eta \rho_0 c^2}{3\omega^4 \rho^2}\left(1 - \omega^2 \rho^2/c^2\right)^{3/2} - \frac{2\pi \eta \rho_0}{\omega^2} + 1 \right] \mathbf{v}_r. \quad (31)$$

For example, consider a neutron star as a relativistic uniform system [21]. Such the star with the mass of 1.35 solar masses and the radius of $a = 12$ km has the average mass density $\rho_0 = 3.7 \times 10^{17}$ kg/m$^3$, and the Lorentz factor at the center can reach the value $\gamma_c = 1.04$. For the star we need to take into account the relation $\eta = \dfrac{3}{5}G$, where $G$ is the gravitational constant, and the angle $\delta_s = \dfrac{a}{c}\sqrt{4\pi \eta \rho_0} \approx 0.546$ radians. Based on these data, the averaged Lorentz factor for the proper motion of particles in (30) turns out to be $\hat{\gamma}' \approx 1.01$, that is less than $\gamma_c$, as it could be expected for the volume-averaged value. This calculation shows that the contribution of $\hat{\gamma}'$ to the vector potential (31) for stars is rather small.

### 4. The pressure field

The equations for the strengths and potentials of the vector pressure field within the framework of the special theory of relativity have the following form [17], [19]:

$$\nabla \cdot \mathbf{C} = 4\pi \sigma \gamma \rho_0, \qquad \nabla \times \mathbf{I} = \frac{1}{c^2}\frac{\partial \mathbf{C}}{\partial t} + \frac{4\pi \sigma \gamma \rho_0 \mathbf{v}}{c^2}, \qquad \nabla \cdot \mathbf{I} = 0, \qquad \nabla \times \mathbf{C} = -\frac{\partial \mathbf{I}}{\partial t}. \qquad (32)$$

$$\partial_\beta \partial^\beta \wp = \frac{1}{c^2}\frac{\partial^2 \wp}{\partial t^2} - \Delta \wp = 4\pi \sigma \rho_0 \gamma, \qquad \partial_\beta \partial^\beta \mathbf{\Pi} = \frac{1}{c^2}\frac{\partial^2 \mathbf{\Pi}}{\partial t^2} - \Delta \mathbf{\Pi} = \frac{4\pi \sigma}{c^2} \mathbf{J}. \qquad (33)$$

$$\mathbf{C} = -\nabla \wp - \frac{\partial \mathbf{\Pi}}{\partial t}, \qquad \mathbf{I} = \nabla \times \mathbf{\Pi}, \qquad \pi_\mu = \left(\frac{\wp}{c}, -\mathbf{\Pi}\right). \qquad (34)$$

Here $\mathbf{C}$ and $\mathbf{I}$ are the strength and the solenoidal vector of the pressure field, respectively, $\sigma$ is the pressure field coefficient, $\pi_\mu$ is the four-potential of the pressure field, $\wp$ and $\mathbf{\Pi}$ are the scalar and vector potentials, $\mathbf{J} = \gamma \rho_0 \mathbf{v}$. Wave equations (33) for the potentials are obtained from equations (32), taking into account (34).

If the system of particles has a spherical shape and rotates at the constant angular velocity $\omega$, the potentials will not depend on the time. Then from (33) it follows:

$$\Delta \wp = -4\pi \sigma \rho_0 \gamma, \qquad \Delta \mathbf{\Pi} = -\frac{4\pi \sigma \rho_0 \gamma}{c^2} \mathbf{v}. \qquad (35)$$

We suppose that in the reference frame $K'$ rotating together with the matter at the angular velocity $\omega$, the Lorentz factor $\gamma'$ has the same form as in (5), corresponding to the system of particles that are moving only randomly in a generally non-rotating system. Substituting $\gamma'$ instead of $\gamma$ into (35), we find the scalar potential of the pressure field for a generally non-rotating system [17]:

$$\Delta \wp' = -4\pi \sigma \rho_0 \gamma'.$$

$$\wp'(\omega=0)=\wp_c-\frac{\sigma c^2\gamma_c}{\eta}+\frac{\sigma c^3\gamma_c}{\eta r\sqrt{4\pi\eta\rho_0}}\sin\left(\frac{r}{c}\sqrt{4\pi\eta\rho_0}\right)\approx\wp_c-\frac{2\pi\sigma\rho_0 r^2\gamma_c}{3}. \qquad (36)$$

If we do not take into account the particles' proper motion at the velocity $\mathbf{v}'$ and the Lorentz factor $\gamma'$, and take into account only the particles' motion due to rotation, in (35) we should substitute the Lorentz factor $\gamma_r$ (6) instead of $\gamma$. This gives the equation for the scalar potential of rotating cylinder in cylindrical coordinates, similar to (11):

$$\Delta\wp_r=-4\pi\sigma\rho_0\gamma_r=-\frac{4\pi\sigma\rho_0}{\sqrt{1-\omega^2\rho^2/c^2}}. \qquad (37)$$

Consequently, the solution of equation (37) will be similar to (14):

$$\wp_r=\frac{4\pi\sigma\rho_0 c^2}{\omega^2}\sqrt{1-\omega^2\rho^2/c^2}-\frac{4\pi\sigma\rho_0 c^2}{\omega^2}\ln\left(1+\sqrt{1-\omega^2\rho^2/c^2}\right)+C_{11}. \qquad (38)$$

For comparison, in [25] for the case of rectilinear motion at the constant velocity of a solid particle without proper rotation, the four-potential of the pressure field was determined as follows:

$$\pi_\mu=\frac{p_0}{\rho_0 c^2}u_\mu=\left(\frac{\wp}{c},-\mathbf{\Pi}\right), \qquad (39)$$

where $p_0$ and $\rho_0$ denote the invariant pressure and mass density in the particle's reference frame $K_p$, the dimensionless ratio $\frac{p_0}{\rho_0 c^2}$ is proportional to the particle's pressure energy per unit of the particle's mass, $u_\mu=(c\gamma,-\gamma\mathbf{v})$ is the particle's four-velocity with a covariant index. In this case, the scalar potential in (39) equals $\wp=\frac{p_0\gamma}{\rho_0}$, and the vector potential equals $\mathbf{\Pi}=\frac{p_0\gamma}{\rho_0 c^2}\mathbf{v}$.

Before determining the constant $C_{11}$ in (38), we will find the solution of equation (35) for a rotating spherical system taking into account the particles' proper random motion. This means that in (35), we must use the expression $\bar{\gamma} = \gamma' \gamma_r$ instead of $\gamma$:

$$\Delta \wp = -4\pi \sigma \rho_0 \gamma' \gamma_r. \tag{40}$$

An approximate solution of equation (40) is similar to the solution of equation (7), and it can be written similarly to (18):

$$\wp \approx \frac{\sigma c^3 \gamma_c}{\eta r \sqrt{4\pi \eta \rho_0}} \left( \frac{1}{\sqrt{1 - \omega^2 \rho^2 / c^2}} + \frac{\omega^2}{2\pi \eta \rho_0} \right) \sin\left( \frac{r}{c} \sqrt{4\pi \eta \rho_0} \right) - \\ - \frac{3 \sigma \gamma_c \omega^4 \rho^4}{8 \eta c^2} + \frac{\pi \sigma \rho_0 \gamma_c \omega^2 \rho^4}{6 c^2} - C_{12}. \tag{41}$$

The constant $C_{12}$ is present in (41), since the potential in (39) is determined with an accuracy up to a constant. For small $r$, we can assume that in (41) $\sin\left( \frac{r}{c} \sqrt{4\pi \eta \rho_0} \right) \approx \frac{r}{c} \sqrt{4\pi \eta \rho_0}$. This gives the following:

$$\wp(r \approx 0) \approx \frac{\sigma c^2 \gamma_c}{\eta} \left( \frac{1}{\sqrt{1 - \omega^2 \rho^2 / c^2}} + \frac{\omega^2}{2\pi \eta \rho_0} \right) - \frac{3 \sigma \gamma_c \omega^4 \rho^4}{8 \eta c^2} + \frac{\pi \sigma \rho_0 \gamma_c \omega^2 \rho^4}{6 c^2} - C_{12}. \tag{42}$$

We can equate expressions (38) and (42) to each other, if in these expressions we direct the cylindrical coordinate $\rho$ to zero. Thus, we will carry out one of the possible options for calibrating the potential of a rotating cylinder. This allows expressing $C_{11}$ and refining the form (38):

$$C_{11} = -\frac{4\pi \sigma \rho_0 c^2}{\omega^2} + \frac{4\pi \sigma \rho_0 c^2}{\omega^2} \ln 2 + \frac{\sigma c^2 \gamma_c}{\eta} \left( 1 + \frac{\omega^2}{2\pi \eta \rho_0} \right) - C_{12}.$$

$$\wp_r = \frac{4\pi \sigma \rho_0 c^2}{\omega^2}\left[\sqrt{1-\omega^2\rho^2/c^2} - \ln\left(1+\sqrt{1-\omega^2\rho^2/c^2}\right) - 1 + \ln 2\right] +$$
$$+ \frac{\sigma c^2 \gamma_c}{\eta}\left(1 + \frac{\omega^2}{2\pi\eta\rho_0}\right) - C_{12}. \tag{43}$$

On the axis of rotation at $\rho = 0$ in (43) will be: $\wp_r(\rho=0) = \frac{\sigma c^2 \gamma_c}{\eta}\left(1 + \frac{\omega^2}{2\pi\eta\rho_0}\right) - C_{12}$. We choose $C_{12}$ so that $\wp_r(\rho=0)$ on the axis of rotation does not depend on the angular velocity $\omega$ of rotation. This is possible if $C_{12} = \frac{\sigma c^2 \gamma_c \omega^2}{2\pi\eta^2\rho_0} - C_{13}$, and $C_{13}$ does not depend on $\omega$. In this case, it will be $\wp_r(\rho=0) = \frac{\sigma c^2 \gamma_c}{\eta} + C_{13}$.

Substitution of $C_{12} = \frac{\sigma c^2 \gamma_c \omega^2}{2\pi\eta^2\rho_0} - C_{13}$ in (41) and (43) gives:

$$\wp \approx \frac{\sigma c^3 \gamma_c}{\eta r\sqrt{4\pi\eta\rho_0}}\left(\frac{1}{\sqrt{1-\omega^2\rho^2/c^2}} + \frac{\omega^2}{2\pi\eta\rho_0}\right)\sin\left(\frac{r}{c}\sqrt{4\pi\eta\rho_0}\right) -$$
$$-\frac{3\sigma\gamma_c\omega^4\rho^4}{8\eta c^2} + \frac{\pi\sigma\rho_0\gamma_c\omega^2\rho^4}{6c^2} - \frac{\sigma c^2 \gamma_c \omega^2}{2\pi\eta^2\rho_0} + C_{13}. \tag{44}$$

$$\wp_r = \frac{4\pi\sigma\rho_0 c^2}{\omega^2}\left[\sqrt{1-\omega^2\rho^2/c^2} - \ln\left(1+\sqrt{1-\omega^2\rho^2/c^2}\right) - 1 + \ln 2\right] + \frac{\sigma c^2\gamma_c}{\eta} + C_{13}. \tag{45}$$

In order to find $C_{13}$, in (44) we can assume the angular velocity of rotation equal to $\omega = 0$ and compare the result with (36) for the case of a non-rotating sphere. This gives the following:

$$C_{13} = \wp_c - \frac{\sigma c^2 \gamma_c}{\eta}. \tag{46}$$

On the other hand, the following expressions were found in [27]:

$$\gamma_c = \frac{1}{\sqrt{1-v_c^2/c^2}} \approx 1 + \frac{3\eta m}{10ac^2}\left(1+\frac{9}{2\sqrt{14}}\right), \quad \wp_c \approx \frac{3\sigma m}{10a}\left(1+\frac{9}{2\sqrt{14}}\right), \quad \frac{\sigma}{\eta} = \frac{2}{3}. \tag{47}$$

Hence it follows that the scalar potential of the pressure field at the center of the sphere is expressed through the root mean square velocity $v_c$ of particles at the center of the sphere:

$$\wp_c = \frac{\sigma c^2 (\gamma_c - 1)}{\eta} = \frac{2c^2 (\gamma_c - 1)}{3} \approx \frac{v_c^2}{3}, \tag{48}$$

Moreover, in (46) will be:

$$C_{13} = -\frac{\sigma c^2}{\eta} = -\frac{2c^2}{3}. \tag{49}$$

Taking into account (46-49) expressions (44-45) will be written as follows:

$$\wp \approx \frac{2c^3 \gamma_c}{3r\sqrt{4\pi \eta \rho_0}} \left( \frac{1}{\sqrt{1 - \omega^2 \rho^2/c^2}} + \frac{\omega^2}{2\pi \eta \rho_0} \right) \sin\left( \frac{r}{c} \sqrt{4\pi \eta \rho_0} \right) -$$
$$-\frac{\gamma_c \omega^4 \rho^4}{4c^2} + \frac{\pi \eta \rho_0 \gamma_c \omega^2 \rho^4}{9c^2} - \frac{c^2 \gamma_c \omega^2}{3\pi \eta \rho_0} - \frac{2c^2}{3}. \tag{50}$$

$$\wp_r = \frac{4\pi \sigma \rho_0 c^2}{\omega^2} \left[ \sqrt{1 - \omega^2 \rho^2/c^2} - \ln\left(1 + \sqrt{1 - \omega^2 \rho^2/c^2}\right) - 1 + \ln 2 \right] + \frac{2c^2 (\gamma_c - 1)}{3}. \tag{51}$$

At small $\omega$, the potential of the pressure field (51) of a rotating cylinder ceases to depend on $\omega$ and becomes approximately equal to $\wp_r \approx \frac{2c^2 (\gamma_c - 1)}{3} - \pi \sigma \rho_0 \rho^2$.

### 5. The vector potential of the pressure field

Substituting in (35) the Lorentz factor $\gamma_r = \frac{1}{\sqrt{1 - \omega^2 \rho^2/c^2}}$ instead of $\gamma$, we obtain the equation for the vector potential of the pressure field for the case of rotation of a cylinder without taking into account the proper motion of the matter's particles:

$$\Delta \mathbf{\Pi}_r = -\frac{4\pi \sigma \rho_0}{c^2 \sqrt{1 - \omega^2 \rho^2/c^2}} \mathbf{v}_r. \tag{52}$$

The velocity $\mathbf{v}_r$ of the particles' motion in (52) is determined in (23). Since (52) coincides in its form with equation (22), then the solution of (52) for the potential component along the axis $OX$ is an expression that repeats (26):

$$\Pi_{rx} = \left[ \frac{4\pi\sigma\rho_0 c^2}{3\omega^3 \rho}\left(1 - \omega^2 \rho^2/c^2\right)^{3/2} - \frac{4\pi\sigma\rho_0 c^2}{3\omega^3 \rho} + \frac{C_{14}}{2}\rho \right]\sin\varphi. \quad (53)$$

According to (39), the scalar potential of the pressure field of a rectilinearly moving particle is equal to $\wp = \dfrac{p_0 \gamma}{\rho_0}$, and the vector potential must be equal to the value $\mathbf{\Pi} = \dfrac{p_0 \gamma}{\rho_0 c^2}\mathbf{v}$. For the scalar potential $\wp_r$ and the component $\Pi_{rx}$ of the vector potential in case of low angular velocities $\omega$ of the cylinder's rotation, in view of (23) and $\gamma_r = \dfrac{1}{\sqrt{1 - \omega^2 \rho^2/c^2}}$ this can be written in the following form:

$$\wp_r \approx \frac{p_0 \gamma_r}{\rho_0} = \frac{p_0}{\rho_0 \sqrt{1 - \omega^2 \rho^2/c^2}}, \qquad \Pi_{rx} \approx \frac{p_0 \gamma_r}{\rho_0 c^2}v_{rx} = -\frac{p_0 \omega \rho \sin\varphi}{\rho_0 c^2 \sqrt{1 - \omega^2 \rho^2/c^2}}.$$

Comparing $\wp_r$ and (51) for small $\omega$ and $\rho$, we find an estimate of the ratio $\dfrac{p_0}{\rho_0}$, which allows us to determine more precisely the expression for $\Pi_{rx}$, in view of (47):

$$\frac{p_0}{\rho_0} \approx \frac{2c^2 \gamma_c}{3} - \frac{2c^2}{3} = \wp_c, \qquad \Pi_{rx} \approx -\frac{\wp_c \omega \rho \sin\varphi}{c^2 \sqrt{1 - \omega^2 \rho^2/c^2}}. \quad (54)$$

Equating $\Pi_{rx}$ in (53) and (54) for low values $\omega$ and $\rho$, we find the constant $C_{14}$ and determine more precisely the expression for $\Pi_{rx}$:

$$C_{14} = \frac{4\pi\sigma\rho_0}{\omega} - \frac{2\wp_c \omega}{c^2}.$$

$$\Pi_{rx} = \left[ \frac{4\pi \sigma \rho_0 c^2}{3\omega^3 \rho}\left(1-\omega^2 \rho^2/c^2\right)^{3/2} - \frac{4\pi \sigma \rho_0 c^2}{3\omega^3 \rho} + \frac{2\pi \sigma \rho_0 \rho}{\omega} - \frac{\wp_c \omega \rho}{c^2} \right] \sin\varphi.$$

Hence, in view of (23), the vector potential of the pressure field of rotating cylinder will be equal to:

$$\mathbf{\Pi}_r = \left[ \frac{4\pi \sigma \rho_0 c^2}{3\omega^4 \rho^2} - \frac{4\pi \sigma \rho_0 c^2}{3\omega^4 \rho^2}\left(1-\omega^2 \rho^2/c^2\right)^{3/2} - \frac{2\pi \sigma \rho_0}{\omega^2} + \frac{\wp_c}{c^2} \right] \mathbf{v}_r. \qquad (55)$$

In this case, the vector potential is directed along the linear velocity $\mathbf{v}_r$, and at low angular velocities $\omega$ it becomes equal to $\mathbf{\Pi}_r \approx \left( \frac{\wp_c}{c^2} - \frac{\pi \sigma \rho_0 \rho^2}{2c^2} \right) \mathbf{v}_r$. Near the axis of rotation, where $\rho \approx 0$, in view of (47), it will be equal to $\mathbf{\Pi}_r \approx \frac{\wp_c}{c^2} \mathbf{v}_r = \left( \frac{2\gamma_c}{3} - \frac{2}{3} \right) \mathbf{v}_r$.

By analogy with (31), we will multiply $\mathbf{\Pi}_r$ by $\hat{\gamma}'$ in order to take into account the contribution of the particles' proper motion to the vector potential of the pressure field of a rotating sphere:

$$\mathbf{\Pi} \approx \hat{\gamma}'\mathbf{\Pi}_r = \hat{\gamma}'\left[ \frac{4\pi \sigma \rho_0 c^2}{3\omega^4 \rho^2} - \frac{4\pi \sigma \rho_0 c^2}{3\omega^4 \rho^2}\left(1-\omega^2 \rho^2/c^2\right)^{3/2} - \frac{2\pi \sigma \rho_0}{\omega^2} + \frac{\wp_c}{c^2} \right] \mathbf{v}_r. \qquad (56)$$

According to (30), the value $\hat{\gamma}'$ is close to unity even for a neutron star.

### 6. The invariant mass of all particles

The total invariant mass of particles of a relativistic uniform system without general rotation is determined by the formula [18]:

$$m_b = \rho_0 \int \gamma'(\omega=0) dV_s = \frac{c^2 \gamma'_c}{\eta}\left[ \frac{c}{\sqrt{4\pi\eta\rho_0}} \sin\left(\frac{a}{c}\sqrt{4\pi\eta\rho_0}\right) - a\cos\left(\frac{a}{c}\sqrt{4\pi\eta\rho_0}\right) \right] \approx$$

$$\approx \frac{4\pi\rho_0 a^3 \gamma'_c}{3}\left(1 - \frac{2\pi\eta\rho_0 a^2}{5c^2}\right).$$

If the system under consideration rotates with angular velocity $\omega$, then to calculate the invariant mass of all particles in the system, the averaged Lorentz factor $\bar{\gamma} = \gamma'\gamma_r$, which is a consequence of (4), should be used. Taking into account (5) and the expression for rotating volume element $dV_s = \dfrac{r^2 dr d\phi \sin\theta d\theta}{\gamma_r}$, we have in spherical coordinates:

$$m_\omega = \rho_0 \int \gamma'\gamma_r dV_s = \frac{c\rho_0 \gamma_c}{\sqrt{4\pi\eta\rho_0}} \int \frac{1}{r} \sin\left(\frac{r}{c}\sqrt{4\pi\eta\rho_0}\right) r^2 dr d\phi \sin\theta d\theta =$$

$$= \frac{c^2 \gamma_c}{\eta}\left[\frac{c}{\sqrt{4\pi\eta\rho_0}}\sin\left(\frac{a}{c}\sqrt{4\pi\eta\rho_0}\right) - a\cos\left(\frac{a}{c}\sqrt{4\pi\eta\rho_0}\right)\right].$$

Since the total invariant mass does not depend on the nature of the particle motion, the masses $m_b$ and $m_\omega$ should be equal to each other. Hence there is the connection between the Lorentz factors at the center of the sphere: $\gamma'_c = \gamma_c$.

This means that in the approximation we have adopted, the particles at the center of the sphere in the rotating frame of reference $K'$ move with the same speeds as in the case of a system of particles without general rotation.

### 7. Conclusion

Wave equations (3) of the acceleration field contain the scalar potential $\vartheta$ and the vector potential $\mathbf{U}$, which depend on the Lorentz factor $\gamma$ and on the velocity $\mathbf{v}$ of the matter's particles. In the general case, the particles move randomly with the Lorentz factor $\gamma'$ (5) and at the same time they participate in collective motion, for example, in general rotation with the Lorentz factor $\gamma_r$ (6). As a result, the Lorentz factor $\gamma$ becomes dependent on $\gamma'$ and $\gamma_r$. Besides, it becomes necessary to average the value $\gamma$ for the system's typical particles in order to use it subsequently in wave equations for the potentials (7-8).

In addition to the scalar potential $\vartheta'$ (10) of the acceleration field for a generally motionless system of particles of spherical shape, we found expressions for the scalar potential in two other cases. One of these cases involves rotation of an infinitely long cylinder, in which the proper motion of the matter's particles is completely neglected. In this case, the scalar potential $\vartheta_r$ is expressed by formula (20). Another case describes rotation of the sphere's matter with angular velocity $\omega$, whereas the system of particles is a relativistic uniform system. The fact that the random proper motion of the particles contributes to the scalar potential $\vartheta$ of a rotating system of particles is reflected in formula (21).

The potential $\vartheta$ in the absence of rotation, when $\omega = 0$, turns into the potential $\vartheta'$ of the fixed system. As for the vector potential $\mathbf{U}$ (31) of the acceleration field, it exceeds the vector potential $\mathbf{U}_r$ (29) of a cylinder by a small coefficient $\hat{\gamma}'$, according to (30).

The considered approach is fully applicable to the pressure field with the exception that the scalar potential of the acceleration field is close in its value to the square of the speed of light, and the scalar potential of the pressure field at the center of the system, according to (48), approximately equals 1/3 of the square of the particle's root mean square velocity.

If we use the definition $\wp = \dfrac{p_0 \gamma}{\rho_0}$ of the scalar potential of pressure field in (39), then for the potential at the center of the sphere we can write: $\wp_c = \dfrac{p_{0c} \gamma_c^2}{\rho_0 \gamma_c} = \dfrac{p_c}{\rho_c}$, where $p_c = p_{0c} \gamma_c^2$ and $\rho_c = \rho_0 \gamma_c$ denote the pressure and density of the matter moving at the center [27]. Comparing this with (48), we arrive at a relativistic formula for the pressure at the center for the case of a non-rotating sphere, that at low velocities transforms into a formula from the molecular kinetic theory: $p_c = \dfrac{2\rho_c c^2 (\gamma_c - 1)}{3} = \dfrac{2\rho_c c^2}{3}\left( \dfrac{1}{\sqrt{1 - v_c^2/c^2}} - 1 \right) \approx \dfrac{\rho_c v_c^2}{3}$. Here $v_c$ is the root-mean-square velocity of typical particles and $\gamma_c = \dfrac{1}{\sqrt{1 - v_c^2/c^2}}$, accordingly, the Lorentz factor at the center of the sphere.

On the other hand, from (10), (36) and (47-48) the relation follows:

$$\wp'(\omega = 0) = \dfrac{\sigma(\vartheta' - c^2)}{\eta} = \dfrac{2c^2 \gamma'(\omega = 0)}{3} - \dfrac{2c^2}{3} \approx \dfrac{v'^2}{3},$$

so that the scalar potential of the pressure field at some point inside the sphere without rotation is three times less than the square of the speed of motion of typical particles at the given point.

From (21) and (50), a similar relation $\wp = \dfrac{\sigma(\vartheta - c^2)}{\eta} = \dfrac{2(\vartheta - c^2)}{3}$ is obtained that connects the potentials of the pressure field and the field of accelerations in the matter of rotating sphere.

Previously, such connections between the potentials of the acceleration field and the pressure field were unknown.

Now, after calculating the scalar and vector potentials of the acceleration field and pressure field, it becomes possible to determine the strengths and solenoidal vectors of these fields by formulas (2) and (34). In turn, the strengths and solenoidal vectors of these fields are required to set the equation of motion of the matter and calculate the relativistic energy of a rotating body, taking into account the energy of not only the particles, but also the fields themselves. Thus, it will be possible to find all the most important parameters of the system. Such work is planned to be carried out in the next article.

As an example, we can take the scalar potential $\vartheta$ and the vector potential $\mathbf{U}$ of the acceleration field, substitute these potentials in (2) and obtain vectors $\mathbf{S}$ and $\mathbf{N}$. The acceleration $\mathbf{a}$ of a typical particle in the limit of the special theory of relativity will be determined by the formula [19]:

$$\mathbf{a} = \mathbf{S} + v \times \mathbf{N} = -\nabla \vartheta - \frac{\partial \mathbf{U}}{\partial t} + \mathbf{v} \times \nabla \times \mathbf{U}, \qquad (57)$$

where $\mathbf{v}$ is the velocity of a particle.

Let us now compare the acceleration $\mathbf{a}$ in (57) with the formula for the acceleration $\boldsymbol{\gamma}$ of a particle presented in [28, Eq. (5)]:

$$\boldsymbol{\gamma} = -\nabla \phi + \nabla \times \boldsymbol{\psi}. \qquad (58)$$

The potential $\vartheta$ coincides in meaning with the potential $\phi$ in (58), however, vector potentials $\mathbf{U}$ and $\boldsymbol{\psi}$ have different meanings. This happens because (57) is obtained in a covariant way from the principle of least action for continuous variables, while the quantities in (58) are presented in the framework of discrete classical mechanics, using the Hodge-Helmholtz decomposition for acceleration.

The standard equation of hydromechanics of an ideal fluid (Euler's equation taking into account mass forces) looks as follows:

$$\frac{d\mathbf{v}}{dt} = \frac{\partial \mathbf{v}}{\partial t} + (\mathbf{v} \cdot \nabla)\mathbf{v} = \frac{\partial \mathbf{v}}{\partial t} + \nabla\left(\frac{v^2}{2}\right) - \mathbf{v} \times [\nabla \times \mathbf{v}] = \mathbf{a}_m - \frac{1}{\rho_m}\nabla p, \qquad (59)$$

where $\mathbf{v}$ is the velocity of liquid particles; $\mathbf{a}_m$ is the acceleration from mass forces acting in the matter; $\rho_m$ is the mass density of the matter; $p$ is the pressure.

If we designate $\nabla \times \mathbf{v} = \mathbf{\Omega}$ and apply to both sides of equation (59) the rotor operation, we get the Friedmann equation, which is often used for rotary motion:

$$\frac{d\mathbf{\Omega}}{dt} - (\mathbf{\Omega} \cdot \nabla)\mathbf{v} + \mathbf{\Omega}\nabla \cdot \mathbf{v} = \nabla \times \mathbf{a}_m + \frac{1}{\rho_m^2}\nabla\rho_m \times \nabla p . \qquad (60)$$

In the simplest case, when the velocity $\mathbf{v}$ lies in planes parallel to the plane $XOY$ of the coordinate system, and the angular velocity of rotation $\mathbf{\omega}$ is directed along the axis $OZ$ and is constant at each point of the system, there will be $\mathbf{\Omega} = 2\mathbf{\omega}$. Equations (59) and (60) imply that there are no viscosity and thermal conductivity in the liquid and the motion occurs adiabatically without a change in entropy. In addition, pressure is considered as a scalar field and therefore participates in the formation of the force only through the gradient in the form of a term $-\frac{1}{\rho_m}\nabla p$. By solving the equations of motion (59) and (60), either the velocity $\mathbf{v}$ is found, or $\mathbf{\Omega} = \nabla \times \mathbf{v}$, if the acceleration $\mathbf{a}_m$, mass density $\rho_m$ and pressure $p$ are known as functions of coordinates and time.

In contrast, a much more general field theory approach gives the following equation of motion taking into account the acceleration field, the vector pressure field, and the dissipation field [15], [19]:

$$\rho_0 a_\beta = \Phi_{\beta\sigma}J^\sigma + F_{\beta\sigma}j^\sigma + f_{\beta\sigma}J^\sigma + h_{\beta\sigma}J^\sigma . \qquad (61)$$

Here $a_\beta = -\frac{1}{\rho_0}u_{\beta\sigma}J^\sigma$ is the four-acceleration, expressed through the tensor $u_{\beta\sigma}$ of the acceleration field; $J^\sigma$ is the mass four-current; $j^\sigma$ is the charge four-current; $\Phi_{\beta\sigma}$, $F_{\beta\sigma}$, $f_{\beta\sigma}$ and $h_{\beta\sigma}$ are tensors of the gravitational field, electromagnetic field, pressure field, and dissipation field, respectively. The indicated tensors are determined through strengths and solenoidal vectors. Thus, the components of the tensor $u_{\beta\sigma}$ are vectors $\mathbf{S}$ and $\mathbf{N}$ in (1-2), found through the potentials of the acceleration field, and the components of the tensor $f_{\beta\sigma}$ are vectors $\mathbf{C}$ and $\mathbf{I}$ in (32-34), found through the potentials of the pressure field. The equation of motion of matter according to (61) in the limit of the special theory of relativity is written as follows:

$$\frac{d}{dt}\left[\gamma\mathbf{v}\left(1+\frac{p_0}{\rho_0 c^2}+\frac{\alpha}{c^2}\right)\right]=\mathbf{a}_m-\frac{1}{\gamma}\nabla\left(\frac{p_0}{\rho_0}\right)-\frac{1}{\gamma}\nabla\alpha. \qquad (62)$$

where $\gamma$ is the Lorentz factor; $\alpha$ is the dissipation function associated with the scalar potential $\varepsilon$ of the dissipation field by the ratio $\varepsilon=\alpha\gamma$; $\mathbf{a}_m$ is the acceleration from the mass gravitational and electromagnetic forces acting in the matter. According to [25], [27], the following can be written: $\frac{p_0}{\rho_0}=\frac{\wp}{\gamma}$, where $\wp$ is the scalar potential of the pressure field.

Comparison of (62) and (59) reveals a difference related to taking relativistic effects into account. So, in (62), the Lorentz factor $\gamma$ is taken into account, as well as the contribution of the invariant pressure $p_0$, invariant mass density $\rho_0$, and the dissipation function $\alpha$ on the left-hand side of the equality. In addition, on the right-hand side of (62) $\rho_0$ is under the gradient sign $\nabla$, while in (59) the value $\rho_m=\gamma\rho_0$ is outside the gradient.

Equation (61) can be written not in terms of tensors, but directly in terms of the four-potentials of the fields [24]:

$$c^2\frac{d\gamma}{dt}=-\frac{d(\psi+\wp)}{dt}-\frac{\rho_{0q}}{\rho_0}\frac{d\varphi}{dt}+\frac{dx^\nu}{dt}\frac{\partial D_\nu}{\partial t}+\frac{\rho_{0q}}{\rho_0}\frac{dx^\nu}{dt}\frac{\partial A_\nu}{\partial t}+\frac{dx^\nu}{dt}\frac{\partial \pi_\nu}{\partial t}, \qquad (63)$$

$$\frac{d(\gamma\mathbf{v})}{dt}=-\frac{d(\mathbf{D}+\mathbf{\Pi})}{dt}-\frac{\rho_{0q}}{\rho_0}\frac{d\mathbf{A}}{dt}-\frac{dx^\nu}{dt}\partial_i D_\nu-\frac{\rho_{0q}}{\rho_0}\frac{dx^\nu}{dt}\partial_i A_\nu-\frac{dx^\nu}{dt}\partial_i \pi_\nu. \qquad (64)$$

Here, $\mathbf{D}$, $\mathbf{\Pi}$ and $\mathbf{A}$ denote the vector potentials of the gravitational field, pressure field and electromagnetic field, respectively; $\rho_{0q}$ is the charge density of an arbitrary particle in the accompanying reference frame; the index $\nu$ runs through the values 0, 1, 2, 3; the index $i$ runs through the values 1, 2, 3; in Cartesian coordinates, $x^\nu=(ct,x,y,z)$ is the four-radius of the particle; $D_\nu=\left(\frac{\psi}{c},-\mathbf{D}\right)$, $\pi_\nu=\left(\frac{\wp}{c},-\mathbf{\Pi}\right)$ and $A_\nu=\left(\frac{\varphi}{c},-\mathbf{A}\right)$ denote the four-potentials of the gravitational field, pressure field and electromagnetic field, respectively; $\psi$, $\wp$ and $\varphi$ there are scalar potentials of the gravitational field, pressure field and electromagnetic field, respectively.

If necessary, other vector fields, for example, the dissipation field, can be taken into account in a similar way in (63-64). Equation (63) describes the energy balance in the system, and equation (64) is the equation of motion of matter. Thus, if the wave equations for the field potentials are solved in the system, similar to equations (33) for the pressure field, then with the help of these potentials the motion of energy and particles in equations (63-64) can be found. This approach has proven itself well, for example, when solving typical problems in electrodynamics.

The references to the formulas for the scalar and vector field potentials obtained in this article are given in Table 1.

**Таблица 1. The index of formulas for field potentials**

|  | Fixed relativistic system | Rotating cylindrical system | Rotating spherical relativistic system |
|---|---|---|---|
| The scalar potential of the acceleration field | $\vartheta'$ (10) | $\vartheta_r$ (20) | $\vartheta$ (21) |
| The vector potential of the acceleration field | $\mathbf{U}' = 0$ | $\mathbf{U}_r$ (29) | $\mathbf{U}$ (31) |
| The scalar potential of the pressure field | $\wp'$ (36) | $\wp_r$ (51) | $\wp$ (50) |
| The vector potential of the pressure field | $\mathbf{\Pi}' = 0$ | $\mathbf{\Pi}_r$ (55) | $\mathbf{\Pi}$ (56) |

**Conflict of interest**

The authors declare that they have no conflict of interest.